\documentclass{elsart}
\usepackage{graphicx}
\usepackage{amssymb}
\newcommand{\ie}{i.e.}
\begin{document}
\runauthor{J{\c{e}}drzejewski and Derzhko}
\begin{frontmatter}

\title{Ground-state properties of multicomponent
Falicov--Kimball-like models I}

\author[Wroclaw,Lodz]{Janusz J{\c{e}}drzejewski\thanksref{email}},
\author[Wroclaw]{Volodymyr Derzhko}

\address[Wroclaw]{Institute of Theoretical Physics,
University of Wroc{\l}aw, pl. Maksa Borna 9, 50--204 Wroc{\l}aw,
Poland}
\address[Lodz]{Department of Theoretical Physics, University of
{\L}{\'{o}}d{\'{z}}, ul. Pomorska 149/153, 90--236
{\L}\'{o}d\'{z}, Poland}

\thanks[email]{Corresponding author: J. J{\c{e}}drzejewski, phone:
+48 71 3759415, fax: +48 71 3214454, e-mail: jjed@ift.uni.wroc.pl}

\begin{abstract}
We consider a classical lattice gas that consists of more than
one ``species'' of particles (like a spin-$\frac{3}{2}$ Ising
model or the atomic limit of the extended Hubbard model), whose
ground-state phase diagram is macroscopically degenerate. This
gas is coupled component-wise and in the Falicov--Kimball-like
manner to a multicomponent free-fermion gas. We show rigorously
that a component-wise coupling of the classical subsystem to the
quantum one orders the classical subsystem so that the macroscopic
degeneracy is removed.
\end{abstract}

\end{frontmatter}

\section{Introduction}
The Hamiltonians of the so-called strongly correlated electron
systems describe in fact multicomponent gases of quasiparticle
excitations in a crystal. The components are labeled, by orbital
and/or spin degrees of freedom, for instance. As a rule,
revealing cooperative phenomena in such systems, like long-range
orders, is a hard task. Then, in the course of studies, one
simplifies the Hamiltonian by discarding some labels of
quasiparticles, setting some coupling constants to zero and
sending other to infinity. Simultaneously, one looses
possibilities of accounting for some physics. But once a
sufficient level of understanding of a simplified model is
reached, a reverse process starts. This might be a brief history
of the Falicov--Kimball model (and only to some extent of the
Hubbard model) studies.

The original model put forward by Falicov and Kimball
\cite{FK,RFK} refers to two groups of quasiparticles excitations,
called localized holes and itinerant electrons, with the members
of each group labeled by band and spin indices. Only a few, out
of many, on-site interactions of these quasiparticles are
retained and the strength of the on-site hole--hole repulsion is
sent to infinity \cite{RFK}. The extreme simplification of this
model, that consists of one band of spinless electrons and of
spinless holes, is known as the {\em spinless Falicov--Kimball
model\/}. In contradistinction to the case of Hubbard model, it
has been possible to demonstrate, without uncontrollable
approximations, that the spinless Falicov--Kimball model exhibits
a staggered long-range order \cite{BS,KL,Lieb}. Subsequent
research extended considerably this result and supplied us with a
fairly broad knowledge of the phase diagram. Naturally, the
strongest results have been obtained at zero temperature (see
\cite{GM,JL} for more details and an extensive list of
references).

Then, one observes a resurgence of interest in effects of
additional degrees of freedom and additional interactions between
localized particles. Most attention has been paid to effects
related to spin degrees of freedom. Thus models, where the
localized and intinerant quasiparticles carry spin-$\frac{1}{2}$
\cite{BFH,BF,CF,FZ,Farkasovsky1,Farkasovsky2} have been
investigated. In \cite{BFH,BF,Farkasovsky2} a finite repulsion
between localized particles with opposite spins has been
included. Quite recently, it was argued that an interesting
physics could be accounted for by multicomponent Falicov--Kimball
models \cite{ZFLC}. As a preliminary step, a kind of
Falicov--Kimball model with $2s+1$ components of itinerant
particles and $2S+1$ components of localized ones was solved
exactly in the limit of infinite dimensions. Concerning
interactions between localized particles, only an on-site
repulsion between different components of localized particles was
included and its strength was sent to infinity, thereby
restricting the space of states available for the system. Unlike
in the quoted above papers on the spin-$\frac{1}{2}$ case, $s$ may
not be equal to $S$.

The {\em multicomponent Falicov--Kimball models\/}, we referred to
above, belong to a class of systems whose generic Hamiltonian is
of the form
\begin{equation}
\label{intro1} H_{\Lambda}= \sum_{\sigma} T_{\sigma} +
\sum_{x,\sigma,\eta} U_{\sigma \eta} n^{e}_{x,\sigma}
n^{i}_{x,\eta} + H_{i} .
\end{equation}
The first sum, over $\sigma$-components of itinerant fermions
(later on called briefly {\em electrons\/}), accounts for their
hopping. The second sum, over lattice sites $x$,
$\sigma$-components of electrons and $\eta$-components of
localized particles (later on called {\em ions\/}), represents the
usual Falicov--Kimball-like coupling between the itinerant and
localized particles. With $n_{x,\sigma}^{e}$, $n_{x,\eta}^{i}$
being the occupation number operators at site $x$ of
$\sigma$-electrons and $\eta$-ions, respectively, it  contributes
the energy $U_{\sigma \eta}$ if and only if the site $x$ is
occupied by a $\sigma$-electron and an $\eta$-ion. The third
term, $H_{i}$, stands for an interaction between ions, diagonal in
their occupation-number representation. Therefore, the system
described by the Hamiltonian $H_{\Lambda}$ can be looked upon as a
multicomponent classical gas (of ions) coupled to a
multicomponent free-fermion gas (of electrons).

In the main body of this paper we study an instance of
(\ref{intro1}) with a {\em component-wise coupling\/} between the
electrons and ions, \ie, \ we set $U_{\sigma \eta} = U_{\sigma}
\delta_{\sigma \eta}$. The component-wise coupling may appear
naturally in situations, where the components differ
significantly in their physical characteristics. Actually, we
make the coupling component-independent, $U_{\sigma} = U$ for all
$\sigma$, and assume equal number of electrons' and ions'
components, for simplicity. If $H_{i}$ involves only finite-range
interactions that admit a representation in terms of a m-potential
\cite{Slawny}, then some aspects of the ground state of the model,
defined by $H_{\Lambda}$, concerned with the distribution of ions
on the underlying lattice, can readily be investigated by the
methods developed in \cite{KL,GJL}.

However, if the itinerant and localized particles are interpreted
as different sorts of electrons, as is the case in the original
paper by Falicov and Kimball, and the coupling between electrons
stands for a screened Coulomb repulsion, then an {\em
ion-independent coupling}, \ie, \ $U_{\sigma \eta} = U_{\sigma}$,
becomes more appropriate. Such a coupling has been studied
rigorously in \cite{BFH,BF} and in the limit of infinite
dimensions in \cite{ZFLC}. The multicomponent Falicov--Kimball
models with ion-independent coupling will be the subject of our
subsequent paper.

In the next section we provide necessary details and properties of
our model with the component-wise coupling, as well as definitions
useful in considerations of ground-state phase diagrams in the
grand-canonical ensemble. Then, in Section 3 we study the
ground-state phase diagram at the hole--particle symmetry point in
the plane of chemical potentials of electrons and ions, for
arbitrary strength $U$ of the electron--ion coupling. After that,
in Section 4 we go beyond the hole--particle symmetry point, but
only in the strong-coupling regime. Finally, in Section 5, we
provide a summary and discussion of the obtained results.

\section{The model and the grand canonical ground states}

In the sequel we study a version of the model (\ref{intro1})
specified as follows. The underlying finite lattice $\Lambda$,
consisting of $|\Lambda|$ sites, is assumed to be a piece of a
$d$-dimensional hypercubic lattice $\Zset^{d}$, such that it can
be partitioned into two congruent sublattices: the even
sublattice, $\Lambda^{e}$, and the odd sublattice, $\Lambda^{o}$.
Each site of $\Lambda$ has $z=2d$ nearest neighbours. Then, the
nearest-neighbour sites of a site in $\Lambda^{e}$ belong to
$\Lambda^{o}$ and vice versa. We shall use the notation $\langle
x,y \rangle \equiv \langle x,y \rangle_{1}$, $\langle x,y
\rangle_{i}$, $i=1,2, \ldots$~, $x,y \in \Lambda$, to denote the
first and the $i$-th nearest neighbours on $\Lambda$, that is the
unordered pairs of sites whose Euclidean distance in the units of
the lattice constant are $1, \sqrt{2}, 2, \ldots$ , for $i=1,2,3,
\ldots $, respectively. Consequently, $\sum_{\langle x,y
\rangle_i}$ will stand for the summation over all the $i$-th
nearest-neighbour pairs of sites, with each pair counted once.

The quantum subsystem, consisting of $r \geqslant 2$ different
kinds of itinerant particles (electrons) labelled by $\sigma =
1,2,\ldots,r$, is described in  terms of creation and annihilation
operators of a $\sigma$-electron at site $x \in \Lambda$:
$c^{+}_{x,\sigma}$, $c_{x,\sigma}$, respectively. The operators
$\left\{ c^{+}_{x,\sigma}, c_{x,\sigma} \right\}^{\sigma =
1,\ldots,r}_{x\in \Lambda}$ satisfy the canonical anticommutation
relations. Then, the hopping of $\sigma$-electrons is specified
by:
\begin{equation}
\label{model1} T_{\sigma}=t\sum_{\langle x,y \rangle} \left(
c_{x,\sigma}^{+} c_{y,\sigma} + {\rm{h.c.}} \right),
\end{equation}
with $t$ being the nearest-neighbour hopping intensity. Another
relevant observable is the total electron-number operator $N_{e}=
\sum_{\sigma} N_{e,\sigma}$, where $\sigma$-electron-number
operators $N_{e,\sigma} = \sum_{x} n^{e}_{x,\sigma}$ and
$n^{e}_{x,\sigma} = c^{+}_{x,\sigma} c_{x,\sigma}$ .

The classical subsystem consists also of $r \geqslant 2$ different
kinds of particles (localized ions). The states of this subsystem
can be described by a collection of particle-occupation numbers $
C = \left\{ n^{i}_{x,\sigma} \right\}^{\sigma=1,\ldots,r}_{x\in
\Lambda}$, with $n^{i}_{x,\sigma} = 0, 1$ (no two ions occupying
the same site can be of the same kind), called the {\em ion
configurations\/}. The total number of ions is
$N_{i}=\sum_{\sigma} N_{i,\sigma}$, where the numbers of
$\sigma$-ions $N_{i,\sigma} = \sum_{x} n^{i}_{x,\sigma}$. In
contradistinction to the electron subsystem, we introduce a
direct ion-ion interaction. Each pair of ions occupying a site
contributes the energy $2I$. Moreover, two ions that occupy two
nearest-neighbour sites repell each other, contributing the
energy $W/2 > 0$. Thus, the ion--ion interaction Hamiltonian,
$H_{i}$, reads
\begin{equation}
\label{model2} H_{i}=2I \sum_{\sigma^{\prime}< \sigma^{\prime
\prime}} \sum_{x} n_{x,\sigma^{\prime}}^{i}
n^{i}_{x,\sigma^{\prime \prime}}+ \frac{W}{2} \sum_{\langle x,y
\rangle} n_{x}^{i} n_{y}^{i} ,
\end{equation}
where $n^{i}_{x} = \sum_{\sigma} n^{i}_{x,\sigma}$ and
$\sum_{\sigma^{\prime}< \sigma^{\prime \prime}}$ stands for the
summation over unordered pairs of indices $(\sigma^{\prime},
\sigma^{\prime \prime})$, each pair being counted once.

Clearly, in the composite system, whose Hamiltonian is given by
(\ref{intro1}), (\ref{model1})  and (\ref{model2}) (with arbitrary
matrix $U_{\sigma \eta}$), the particle-number operators
$N_{e,\sigma}$ (hence $N_{e}$) are conserved. The peculiarity of
this system is that the site-occupation numbers
$n^{i}_{x,\sigma}$ (hence $N_{i,\sigma}$ and $N_{i}$)  are also
conserved  and therefore the description of the classical
subsystem in terms of the ion configurations $C =\left\{
n_{x,\sigma}^{i} \right\}^{\sigma=1,\ldots,r}_{x \in \Lambda}$
remains valid.

Finally, the coupling between the classical and quantum
subsystems is chosen to be the component-wise one, with
\begin{equation}
\label{model3} U_{\sigma \eta} = 2U \delta_{\sigma \eta} .
\end{equation}
This concludes the specification of our system.

In what follows, we shall study the ground-state phase diagram of
the system defined by (\ref{intro1})--(\ref{model3}), in the
grand canonical ensemble. Ultimately, this task amounts to
determining at each point $\left(\tilde{\mu}_{e}, \tilde{\mu}_{i}
\right )$ of the plane of chemical potentials of electrons and
ions, respectively, the lowest energy eigensubspace of the
Hamiltonian
\begin{equation}
\label{model4} \tilde{H}_{\Lambda} \left( \tilde{\mu}_{e},
\tilde{\mu}_{i}
\right)=H_{\Lambda}-\tilde{\mu}_{e}N_{e}-\tilde{\mu}_{i}N_{i} .
\end{equation}
Actually, our purposes will be more restricted. Namely, we shall
be interested only in determining the set of periodic ground-state
configurations of ions at each point in the $\left(
\tilde{\mu}_{e},\tilde{\mu}_{i} \right)$ plane.

In studies of grand-canonical phase diagrams an important role is
played by unitary transformations that exchange particles and
holes: $n^{e}_{x,\sigma} \rightarrow 1 - n^{e}_{x,\sigma}$ and
$n^{i}_{x,\sigma} \rightarrow 1 - n^{i}_{x,\sigma}$, and for some
$\left( \tilde{\mu}^{0}_{e},\tilde{\mu}^{0}_{i} \right)$ leave
the Hamiltonian $\tilde{H}_{\Lambda} \left( \tilde{\mu}_{e},
\tilde{\mu}_{i} \right)$ invariant. For the electron subsystem of
(\ref{model4}) such a role is played by the transformations:
$c_{x,\sigma}^{+} \rightarrow (-1)^{|x|}c_{x,\sigma}$, with
$|x|=|(x_{1},\ldots,x_{d})|=\sum_{i=1}^{d} |x_{i}|$. At the
hole--particle symmetry point, $\left( \tilde{\mu}_{e}^{0}
,\tilde{\mu}_{i}^{0} \right)$, the system under consideration has
very special properties, which facilitate studies of its phase
diagram (among them is the property that the grand-canonical
expectations of $n_{x, \sigma}^{e}, n_{x, \sigma}^{i}$ assume the
value of $\frac{1}{2}$, independently of temperature). One easily
finds that this system is hole--particle invariant at the point
$\left( \tilde{\mu}_{e}^{0}, \tilde{\mu}_{i}^{0} \right)$, where
\[
\tilde{\mu}_{e}^{0}=U, \ \ \ \ \tilde{\mu}_{i}^{0}= \tilde{
\nu}_{i}^{0}+U, \ \ \ \ \tilde{\nu}_{i}^{0}= \frac{W}{4}rz+I(r-1).
\]
For future purposes, it is convenient to introduce the shifted
chemical potentials
\[
\mu_{e} = \tilde{\mu}_{e} - \tilde{\mu}_{e}^{0}, \ \ \ \ \mu_{i}
= \tilde{\mu}_{i} - \tilde{\mu}_{i}^{0},
\]
the new Hamiltonian $H_{\Lambda} \left( \mu_{e}, \mu_{i} \right) =
\tilde{H}_{\Lambda} \left( \tilde{\mu}_{e}, \tilde{\mu}_{i}
\right)$, and express $H_{\Lambda} \left( \mu_{e}, \mu_{i}
\right)$ as follows
\begin{equation}
\label{model5-1} H_{\Lambda} \left( \mu_{e}, \mu_{i} \right) =
\sum_{\sigma} \left( H_{\sigma}^{FK} - \mu_{e}N_{e,\sigma}-
\mu_{i}N_{i,\sigma} \right) +H_{i}^{0},
\end{equation}
\begin{eqnarray}
\label{model5-2} H_{\sigma}^{FK} & = & T_{\sigma}+2U\sum_{x}
n_{x,\sigma}^{e}n_{x,\sigma}^{i} - U N_{e,\sigma}- U N_{i,\sigma}
\nonumber \\
& = & T_{\sigma}+2U\sum_{x} \left( n_{x,\sigma}^{e} - \frac{1}{2}
\right)\left( n_{x,\sigma}^{i} - \frac{1}{2} \right),
\end{eqnarray}
\begin{eqnarray}
\label{model5-3} H_{i}^{0} & = & H_{i}-\tilde{\nu}_{i}^{0} N_{i}
\nonumber \\
& = & 2I \sum_{x,{\sigma^{\prime}}<{\sigma^{\prime \prime}}}
\left( n_{x,\sigma^{\prime}}^{i} - \frac{1}{2} \right)\left(
n_{x,\sigma^{\prime \prime}}^{i} - \frac{1}{2} \right)
+\frac{W}{2} \sum_{\langle x,y \rangle} \left( n_{x}^{i} -
\frac{r}{2} \right)\left( n_{y}^{i} - \frac{r}{2} \right).
\end{eqnarray}

The second equalities in (\ref{model5-2}, \ref{model5-3}) hold up
to a term which is independent of the particle states. The
Hamiltonians $H^{FK}_{\sigma}$ and $H^{0}_{i}$ are invariant with
respect to the hole--particle transformation, consequently
$H_{\Lambda} \left( \mu_{e}, \mu_{i} \right)$ is hole--particle
invariant at the point $\mu_{e} = \mu_{i} = 0$.

We define $E_{C} \left( \mu_{e}, \mu_{i} \right)$ to be the
ground-state energy of $H_{\Lambda} \left( \mu_{e}, \mu_{i}
\right)$ for a given configuration of ions $C$. Then the
ground-state energy of the Hamiltonian $H_{\Lambda} \left(
\mu_{e}, \mu_{i} \right)$, denoted $E_{G}\left( \mu_{e}, \mu_{i}
\right)$, is  $E_{G}\left( \mu_{e}, \mu_{i} \right) = \min \left\{
E_{C} \left( \mu_{e}, \mu_{i} \right): C \right\}$, and the
minimum is attained at the set $G$ of the ground-state
configurations of ions.

\section{Ground-state phase diagram at the hole--particle symmetry point}

In this section we shall describe the ground-state configurations
of ions in the systems given by $H^{0}_{i}$ and $H_{\Lambda}
\left( 0,0 \right)$.

Let us consider first the classical subsystem of ions, given by
$H^{0}_{i}$. The nature of its ground-state configurations of ions
is explained by the Theorem \ref{thm1}, which follows. Briefly,
there are two domains of couplings $I$, $W$, differentiated by
the ratio $\delta = 4I/zW$. For $\delta < 1$, there are exactly
two ground-state configurations, while for $\delta \geqslant 1$
there is a macroscopic degeneracy (\ie, \ the number of
ground-state configurations grows exponentially with the size
$|\Lambda |$ of the underlying lattice).

\begin{thm}
\label{thm1} Consider the system given by
$H^{0}_{i}$, with $W>0$. \\
(i) If $4I<zW$, the set of ground-state configurations consists
of exactly two ion configurations with the following property:
each site of one of the sublattices of $\Lambda$ is occupied by
$r$ ions while the complementary sublattice is empty.\\
(ii) If $4I>zW$, the set of ground-state configurations contains
macroscopically many elements.\\
For even $r$, $r=2k$, any ion configuration with $k$ ions in each
site is the ground-state one; the degeneracy of the ground state
is $(C^{k}_{r})^{|\Lambda|}$, where $C_{r}^{k}=r!/k!(r-k)!$. \\
For odd $r$, $r=2k+1$, any ion configuration satisfying the
condition: each site of one sublattice is occupied by $k$ ions
while each site of the complementary sublattice by $(k+1)$ ions,
is the ground-state configuration; the degeneracy is
$(C_{r}^{k})^{|\Lambda|/2} (C_r^{(k+1)})^{|\Lambda|/2}$.\\
(iii) If $4I=zW$ the set of ground-state configurations contains
macroscopically many elements: any configuration of kind (i) or
(ii) is the ground-state configuration.
\end{thm}
A proof of Theorem \ref{thm1} is given in appendix.

Now, consider the composite system that consists of $r$
components of classical ions coupled to $r$ components of quantum
electrons, described by the Hamiltonian $H_{\Lambda}(0,0)$. The
following theorem characterizes the ground-state configurations
of ions at the hole--particle symmetry point.
\begin{thm}
\label{thm2} Consider the system given by $H_{\Lambda}(0,0)$
with $U \neq 0$ and $W>0$.\\
(i) If $4I<zW$, then the ground-state energy is attained for
exactly two ion configurations that satisfy the following
condition. The ions fill out completely one sublattice of
$\Lambda$, \ie, \ each site of this sublattice is occupied by $r$
ions, while the complementary sublattice is empty.\\
(ii) If $4I>zW$, then the set of ground-state configurations of
ions contains finitely many elements; any two nearest-neighbour
sites cannot be occupied by ions of the same kind.\\
Moreover, for $r$ even, $r=2k$, there are exactly $C^{k}_{r}$
ground-state configurations of ions; each site of $\Lambda$ is
occupied by $k$ ions.\\
For $r$ odd, $r=2k+1$, there are exactly $2C^{k}_{r}$ ground-state
configurations of ions with the following property: each site of
one sublattice is occupied by $k$ ions, while each site of the
complementary sublattice contains $k+1$ ions.\\
(iii) If $4I=zW$, the set of ground state configurations of ions
consists of finitely many elements: any configuration of kind (i)
or (ii) is the ground-state configuration.
\end{thm}

Thus, the effect of the component-wise coupling of our classical
gas of ions with quantum electrons amounts to removing the
macroscopic degeneracy of the domain of $I$, $W$, where $\delta
\geqslant 1$. The domain, where $\delta<1$, of finite degeneracy
is left unchanged. This effect does not depend neither on the
hopping intensity $t$ nor on the strength $U$ of the ion-electron
coupling.

To prove the Theorem \ref{thm2}, let us note that since the
operators $H^{FK}_{\sigma}$ ($\sigma = 1, \ldots, r$),
$H^{0}_{i}$, commute pairwise, for given configuration $C$
\begin{equation}
\label{energy-C}
E_{C}(0,0) = \sum_{\sigma} E^{FK}_{C_{\sigma}} + E^{i}_{C},
\end{equation}
where $E^{FK}_{C_{\sigma}}$ is the ground-state energy of
$H^{FK}_{\sigma}$ for the configuration $C_{\sigma} = \left\{
n^{i}_{x,\sigma}\right\}_{x\in \Lambda}$ that is the restriction
of $C$ to $\sigma$-ions, and $E^{i}_{C}$ is the value of
$H^{0}_{i}$ at $C$. Let $E^{FK}_{G}$ be the ground-state energy of
$H^{FK}_{\sigma}$, \ie\ $E^{FK}_{G} = \min \left\{
E^{FK}_{C_{\sigma}}: C_{\sigma}\right\}$ and the minimum is
attatined at the configurations $C_{\sigma}\in G^{FK}_{\sigma}$.
The energy $E^{FK}_{G}$ and the set $G^{FK}_{\sigma}$ do not
depend on $\sigma$. Let $E^{i}_{G} = \min \left\{ E^{i}_{C}: C
\right\}$ and let $G^{i}$ be the set of configurations for which
the minimum is attained. In general
\begin{equation}
\label{energy-G}
E_{G}(0,0) \geq \sum_{\sigma} E^{FK}_{G} + E^{i}_{G}.
\end{equation}
However, we shall show that the lower bound (\ref{energy-G}) of
$E^{FK}_{G}$ is attained for the set G of the configurations
described in Theorem \ref{thm2}.

Consider first the term $\sum_{\sigma} E^{FK}_{C_{\sigma}}$ of
(\ref{energy-C}). Clearly,
\[
\min \left\{ \sum_{\sigma} E^{FK}_{C_{\sigma}}: C=(C_{1}, \ldots,
C_{r}) \right\} = rE^{FK}_{G}
\]
and is attained for configurations in
\[
G^{FK} = \left\{(C_{1}, \ldots, C_{r}): C_{\sigma} \in
G^{FK}_{\sigma}, \sigma = 1,
\ldots, r \right\}.
\]
In order to determine $G^{FK}_{\sigma}$ (and $E^{FK}_{G}$) we
follow \cite{KL}. Note that, for a fixed configuration $C$ the
operator $H^{FK}_{\sigma} + U \sum_{x} \left( n^{i}_{x,\sigma} -
\frac{1}{2} \right)$ is the second quantized form of the
one-particle operator $h_{\sigma} = T + U S_{\sigma}$, where $T$
and $S_{\sigma}$ are $|\Lambda| \times |\Lambda|$ matrices with
the following matrix elements $[T]_{xy}$, $[S_{\sigma}]_{xy}$:
\[
[T]_{xy}= \left\{
\begin{array}{cc}
t, & \mbox{if $x,y$ are nearest neighbours on $\Lambda$} \\
0, & \mbox{otherwise}
\end{array}
\right.
\]
\[
[S_{\sigma}]_{xy}=s_{x,\sigma} \delta_{xy}, \ \ \ \
s_{x,\sigma}=2n_{x,\sigma}^{i}-1.
\]

The lowest eigenvalue of
$H^{FK}_{\sigma}+U\sum_{x}\left(n_{x,\sigma}^{i}-\frac{1}{2}
\right)$ is the Fermi-sea energy of $h_{\sigma}$ corresponding to
the Fermi level $\mu_{e} = 0$. Thus, in terms of the eigenvalues
$\lambda^{\sigma}_{j}$, $j = 1, \ldots, |\Lambda|$, of the
operator $h_{\sigma}$
\begin{eqnarray*}
E_{C_{\sigma}}^{FK}&=&\sum_{\lambda_{j}^{\sigma}<0}
\lambda_{j}^{\sigma} - \frac{U}{2} \sum_{x} s_{x, \sigma} =
-\frac{1}{2} \left[\sum_{\lambda_{j}^{\sigma}} |
\lambda_{j}^{\sigma}|-\sum_{\lambda_{j}^{\sigma}}
\lambda_{j}^{\sigma} \right] - \frac{U}{2} \sum_{x} s_{x, \sigma}
\\
&=& -\frac{1}{2} \left[ {\rm{Tr}}|h_{\sigma}| -
{\rm{Tr}}h_{\sigma} \right]- \frac{U}{2} \sum_{x} s_{x, \sigma} =
-\frac{1}{2} {\rm{Tr}} |h_{\sigma}|,
\end{eqnarray*}
since ${\rm{Tr}} \, T=0$. Kennedy and Lieb have proved in
\cite{KL} that
\[
- {\rm{Tr}} |h_{\sigma}| \geqslant  - {\rm{Tr}} \left(T^{2} +
U^{2} \right)
\]
and the lower bound is attained for exactly two configurations
$C_{\sigma}$ with the following property: one sublattice of
$\Lambda$ is completely occupied by $\sigma$-ions while the
complementary sublattice is empty. Thus, $E^{FK}_{G} = -
\frac{1}{2} {\rm{Tr}} \left(T^{2} + U^{2} \right)$ and the
described above configurations constitute $G^{FK}_{\sigma}$.

The set $G^{i}$ is specified in Theorem \ref{thm1}. Now, since
$G^{FK} \cap G^{i}$ is nonvoid, the set $G$ of ground-state
configurations of $H_{\Lambda}(0,0)$ is $G= G^{FK} \cap G^{i}$ and
the ground-state energy equals to $E_{G}(0,0)= - \frac{r}{2}
{\rm{Tr}} \left(T^{2} + U^{2} \right) + E^{i}_{G}$.

In the following section we shall show that the effect of
removing macroscopic degeneracies of classical-ion phases persists
off the hole--particle symmetry point.

\section{Ground states off the symmetry point in the strong-coupling regime}

Let us consider the system given by (\ref{model5-1}),
(\ref{model5-2}) and (\ref{model5-3}) with $\mu_{i}\neq 0$. In
order to simplify the arguments that follow, we set $r=2$, with
the label $\sigma$ interpreted as spin-$\frac{1}{2}$ and taking
values in $\left\{ \uparrow, \downarrow \right\}$. Moreover, the
calculations are carried out for $d=2$, \ie \ $\Lambda$ is a
piece of the square lattice, but the obtained results can be
generalized to higher-dimensional hypercubic lattices in a
straightforward way.

As in the previous section, we start with the classical subsystem
whose Hamiltonian is $H_{i}^{0}-\mu_{i} N_{i}$, and is specified
by (\ref{model5-3}). The ground-state phase diagram in the plane
$\left( I/W, \mu_{i}/W \right)$ can be obtained by extending the
calculations of Appendix A, and is shown in Fig. 1.

\begin{figure}
\centering \includegraphics[height=7cm]{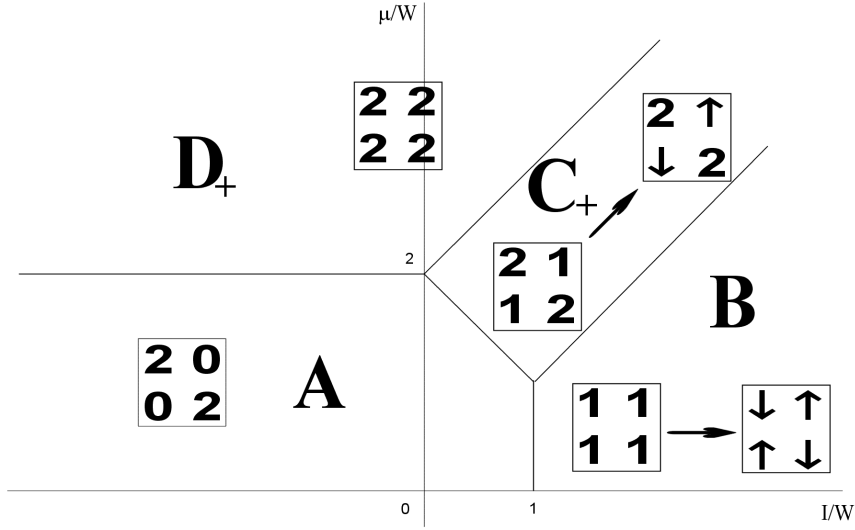}
\caption{Ground-state phase diagram of $H_{i}^{0}-\mu N_{i}$, for
$\mu \geqslant 0$. The ground-state configurations inside the
regions $A$, $B$, $C_{+}$, $D_{+}$ are obtained by periodizing the
elementary-plaquette configurations displayed inside the regions
$A$, $D_{+}$ and those located by the starting points of the
arrows in $B$, $C_{+}$. The elementary-plaquette configurations
located by the end points of the arrows in $B$, $C_{+}$ show the
effect of the coupling to the electron subsystem. The second half
of the diagram, for $\mu <0$, containing the second halves of the
regions $A$, $B$ and the regions $C_{-}$, $D_{-}$, can be
obtained by reflecting the part with $\mu >0$ in the line $\mu=0$
and simultaneously replacing the elementary-plaquette
configurations, by their hole--particle counterparts ($n_{x}^{i}
\rightarrow 2-n_{x}^{i}$).}
\end{figure}

The energies of ion configurations depend only on the occupation
numbers of sites: $n^{i}_{x}=2, 1, 0$. Consequently, the
ground-state configurations in the domains $B$, $C_{+}$, $C_{-}$
are macroscopically degenerate due to the spin degeneracy.

According to the Theorem {\ref{thm2}}, at the line $\mu_{i}=0$ and
in the domain $B$, the coupling to the electron subsystem orders
the ion subsystem antiferromagnetically: each sublattice is
occupied exclusively by one kind of ions.

In order to determine the effect on the macroscopically
degenerate ground-state configurations of the coupling to the
electron subsystem, off the line $\mu_{i}=0$, we resort to the
strong-coupling limit. In this limit, the two components
$H_{\sigma}^{FK}-\mu_{e}N_{e, \sigma}$ of $H_{\Lambda} \left(
\mu_{e}, \mu_{i} \right)$ can be expanded into a power series in
$1/|U|$ \cite{GJL}. The obtained series is convergent absolutely
and uniformly in $|\Lambda|$, provided $|U|>zt$ and
$|\mu_{e}|<|U|-zt$ (here $z=4$) \cite{Kennedy,GMMU}. Due to
symmetries of $H_{\Lambda}\left( \mu_{e}, \mu_{i} \right)$ we can
restrict our considerations to the case $U>0$, and set
$N_{e,\sigma}=|\Lambda|-N_{i, \sigma}$ \cite{KL,GJL}.

Let $H_{\Lambda}^{(k)} \left( \mu_{e}, \mu_{i} \right)$,
$k=1,2,\ldots$, stand for the partial sums of the $1/U$ expansion
of $H_{\Lambda} \left( \mu_{e}, \mu_{i} \right)$. Then, up to a
term independent of the ion configurations, the partial sum
$H_{\Lambda}^{(3)} \left( \mu_{e}, \mu_{i} \right)$ reads
\cite{GJL}:
\begin{eqnarray}
\label{expansion-1} H_{\Lambda}^{(3)}\left( \mu_{e}, \mu_{i}
\right)& & =-\frac{\mu_{i}}{2} \sum\limits_{x} \left( s_{x,
\uparrow}+s_{x, \downarrow} \right)+
\frac{I}{2}\sum\limits_{x}s_{x, \uparrow}s_{x,
\downarrow} \nonumber \\
& &+\frac{W}{8}\sum\limits_{\langle x,y \rangle_{1}} \left( s_{x,
\uparrow} + s_{x, \downarrow} \right) \left( s_{y, \uparrow} +
s_{y,\downarrow}\right) \nonumber \\
& &+\left( \frac{t^{2}}{4U}-\frac{9t^{4}}{16U^{3}}\right)
\sum\limits_{\langle x,y \rangle_{1}} \left( s_{x, \uparrow}
s_{y, \uparrow}+s_{x, \downarrow} s_{y, \downarrow}
\right) \nonumber \\
& &+ \frac{3t^{4}}{16U^{3}}\sum\limits_{\langle x,y \rangle_{2}}
\left( s_{x, \uparrow} s_{y, \uparrow}+s_{x, \downarrow} s_{y,
\downarrow} \right)+\frac{t^{4}}{8U^{3}}\sum\limits_{\langle x,y
\rangle_{3}} \left( s_{x, \uparrow} s_{y, \uparrow} + s_{x,
\downarrow}
s_{y, \downarrow} \right) \nonumber \\
& & +\frac{5t^{4}}{16U^{3}} \sum\limits_{P}\left(s_{x,\uparrow}
s_{y,\uparrow} s_{z,\uparrow} s_{w,\uparrow} + s_{x,\downarrow}
s_{y,\downarrow} s_{z,\downarrow} s_{w,\downarrow}\right)
+O\left(\frac{t^{6}}{U^{5}}\right).
\end{eqnarray}

Since $H_{\Lambda}^{(0)} \left( \mu_{e}, \mu_{i} \right)$ differs
from $H_{i}^{0}-\mu_{i}N_{i}$ by a constant, at the zero order
the phase diagram coincides with that of the classical subsystem.
In the first order, $H_{\Lambda}^{(0)}\left( \mu_{e},\mu_{i}
\right)$ is augmented by the Ising antiferromagnetic interaction
of the strength $t^{2}/4U$. Consequently, in the phase diagram of
$H_{\Lambda}^{(1)} \left( \mu_{e}, \mu_{i} \right)$ the domains
$B$, $C_{+}$, $C_{-}$ are replaced by some domains $B^{(1)}$,
$C_{+}^{(1)}$, $C_{-}^{(1)}$, whose boundaries are shifted with
respect to the boundaries of $B$, $C_{+}$, $C_{-}$ by terms of
the order $1/U$. The set of ground-state configurations inside
$B^{(1)}$ consists only of the two antiferromagnetic
configurations, while the sets of ground-state configurations
inside $C_{+}^{(1)}$, $C_{-}^{(1)}$ remain the same as those in
$C_{+}$, $C_{-}$. Passing to the next order, one finds that the
phase diagram of $H_{\Lambda}^{(3)} \left( \mu_{e}, \mu_{i}
\right)$ contains some domains $B^{(3)}$, $C_{+}^{(3)}$,
$C_{-}^{(3)}$ whose boundaries are shifted with respect to the
boundaries of domains $B^{(1)}$, $C_{+}^{(1)}$, $C_{-}^{(1)}$ by
terms of the order $1/U^{3}$. In this order, the set of
ground-state configurations inside $B^{(3)}$ remains the same as
that in $B^{(1)}$. On the other hand, the macroscopically
degenerate sets of ground-state configurations in $C_{+}^{(1)}$
($C_{-}^{(1)}$) are replaced by exactly four configurations in
$C_{+}^{(3)}$ ($C_{-}^{(3)}$): one sublattice is doubly occupied
(empty) while the complementary sublattice is ordered
antiferromagnetically. Applying the arguments due to Kennedy
\cite{Kennedy} and Gruber et al. \cite{GMMU}, we conclude that
there are nonvoid domains $B^{(\infty)}$, $C_{+}^{(\infty)}$,
$C_{-}^{(\infty)}$ such that $B^{(\infty)}\subset B^{(3)}$,
$C_{+}^{(\infty)}\subset C_{+}^{(3)}$, $C_{-}^{(\infty)}\subset
C_{-}^{(3)}$, inside which the sets of ground-state
configurations of $H_{\Lambda}\left( \mu_{e}, \mu_{i} \right)$
coincide with those inside $B^{(3)}$, $C_{+}^{(3)}$,
$C_{-}^{(3)}$, respectively.

\section{Discussion}

The two-component classical system considered in Section 4 and
given by the Hamiltonian $H_{i}^{0}-\mu_{i}N_{i}$ is well known
in the literature of strongly correlated electron systems as the
atomic limit of the extended Hubbard model. A lot of effort has
been devoted to studies of its phase diagram, see for instance
\cite{MRC,BJK} and references quoted there. When the two kinds of
ions involved (in the context of Hubbard-like models they are
interpreted as localized electrons with spin up or down) are
allowed to hop, we arrive at the extended Hubbard model. This
model is, since many years, a subject of vigorous research. One
of the intriguing questions asked is what is the effect of
hopping on the phases of localized electrons (determined in the
atomic limit). It is widely known that the second-order
perturbation theory, with respect to hopping, predicts that the
ground-state configurations with only singly occupied sites
(inside domain $B$) are ordered antiferromagnetically by a weak
hopping. Brandt and Stolze have shown \cite{BS1} that the
prediction for the phase with density $\frac{3}{2}$
($\frac{1}{2}$), inside domain $C_{+}$ ($C_{-}$), is that it is
ordered ferromagnetically by a weak hopping, what results in
coexistence of charge and magnetic orders. Thus, according to the
second-order perturbation theory the quantum fluctuations,
introduced by the hopping, remove the spin degeneracy present in
the ground-state phases of localized electrons. In this paper we
address an analogous problem. In our case however, the direct
quantum fluctuations introduced by the hopping of the ions are
replaced by the indirect ones, introduced by the hopping of an
extra quantum system of free electrons coupled to the classical
system of ions in a Falicov--Kimball way. In this new set up, the
effect of quantum fluctuations due to the hopping could have been
studied rigorously. In Section 3, we have considered only the
hole--particle symmetry point. For any strength of the coupling
between electrons and ions and any hopping, we obtained that the
ground-state configurations of singly occupied sites (when there
is no hopping of electrons), become antiferromagnetically ordered
upon switching the hopping. This result agrees with the
predictions of the perturbation theory for the extended Hubbard
model. However, far away of the symmetry point, inside the
domains $C_{+}$ ($C_{-}$), where without hopping the ground-state
configurations are spin degenerate, we also find an
antiferromagnetic order upon switching the hopping. This result
is in contrast with the predictions of the perturbation theory
for the extended Hubbard model.

\ack One of the authors (V.D.) is grateful to the University of
Wroc{\l}aw, and especially to the Institute of Theoretical
Physics for financial support.

\appendix
\section*{Appendix}
In order to prove Theorem \ref{thm1}, it is enough to notice that
if $Q_{x}= n^{i}_{x}-r/2$, then (\ref{model5-3}) reads:
\[
H^{0}_{i} = \frac{W}{4} \sum\limits_{\langle x, y \rangle} \left(
Q_{x} + Q_{y}\right)^{2} - \left( \frac{W}{4}z-I \right)
\sum\limits_{x} Q^{2}_{x}.
\]
Rewriting $H^{0}_{i}$ in terms of a nearest-neighbour bond
potential $\phi_{\langle x, y \rangle}$
\[
H^{0}_{i}=\sum\limits_{\langle x, y \rangle} \phi_{\langle x, y
\rangle},
\]
\[
\phi_{\langle x, y \rangle}=\frac{W}{4}(Q_{x}+Q_{y})^{2} - \left(
\frac{W}{4}-\frac{I}{z} \right) \left( Q_{x}^{2}+Q_{y}^{2}
\right),
\]
and minimizing $\phi_{\langle x, y \rangle}$ one finds that it is
a m-potential, which gives us the periodic ground-state
configurations of $H^{0}_{i}$. The set $G^{i}$ of configurations
which realize the minimum of the energy $E_{C}^{i}$ ($W>0$), in
the terms of $Q_{x}$, consists of configurations whose
restriction to the nearest-neighbour bonds $\langle x,y \rangle$
is as follows:
\begin{eqnarray*}
{\mbox{a)}} \,\,\,\, & r=2k, \,\, 4I>zW, & \,\, (Q_{x},Q_{y})=(0,0), \\
{\mbox{b)}} \,\,\,\, & r=2k, \,\, 4I<zW, & \,\, (Q_{x},Q_{y})=(-k,k), \\
{\mbox{c)}} \,\,\,\, & r=2k+1, \,\, 4I>zW, & \,\, (Q_{x},Q_{y})=(-1/2,1/2), \\
{\mbox{d)}} \,\,\,\, & r=2k+1, \,\, 4I<zW, & \,\,
(Q_{x},Q_{y})=(-k-1/2,k+1/2);
\end{eqnarray*}
or in terms of $n_{x}^{i}$
\begin{eqnarray*}
\mbox{a)} \,\,\,\, 4I>zW, & \,\, r=2k+1, & \,\, (n^{i}_{x},n^{i}_{y})=(k,k+1), \\
\mbox{b)} \,\,\,\, 4I>zW, & \,\, r=2k, & \,\, (n^{i}_{x},n^{i}_{y})=(k,k) \\
\mbox{c)} \,\,\,\, 4I<zW, & \,\, & \,\, (n^{i}_{x},n^{i}_{y})=(r,0). \\
\end{eqnarray*}
Periodizing these bond configurations one obtains $G^{i}$.

\end{document}